\documentclass[useAMS,usenatbib]{mnras}
\usepackage{graphicx}
\usepackage{amsmath}
\usepackage{array}
\usepackage{gensymb}
\usepackage{subcaption}
\usepackage{tabularx}

\title[Multiple-population dynamics in NGC 6362]{The dynamics of multiple populations in the globular cluster NGC 6362}
\author[Meghan Miholics, Jeremy J. Webb, Alison Sills]{Meghan Miholics\thanks{E-mail: miholim@mcmaster.ca (MM)}, Jeremy J. Webb, Alison Sills\\
Department of Physics and Astronomy, McMaster University, Hamilton ON, L8S 4M1,Canada}
\begin{document}

%\date{Accepted 1988 December 15. Received 1988 December 14; in original form 1988 October 11}

\pagerange{\pageref{firstpage}--\pageref{lastpage}} \pubyear{2015}

\maketitle

\label{firstpage}

\begin{abstract}

We investigate how the Milky Way tidal field can affect the spatial mixing of multiple stellar populations in the globular cluster NGC 6362. We use $N$-body simulations of multiple population clusters on the orbit of this cluster around the Milky Way. Models of the formation of multiple populations in globular clusters predict that the second population should initially be more centrally concentrated than the first. However, NGC 6362 is comprised of two chemically distinct stellar populations having the same radial distribution. We show that the high mass loss rate experienced on this cluster's orbit significantly accelerates the spatial mixing of the two populations expected from two body relaxation. We also find that for a range of initial second population concentrations, cluster masses, tidal filling factors and fraction of first population stars, a cluster with two populations should be mixed when it has lost 70-80 per cent of its initial mass. These results fully account for the complete spatial mixing of NGC 6362, since, based on its shallow present day mass function, independent studies estimate that the cluster has lost 85 per cent of its initial mass.
\end{abstract}


\section{Introduction}

Globular clusters are dense stellar systems which were believed to be the classic examples of a single stellar population, consisting of stars born at the same time with the same chemical composition. However, observed variations in star to star light element abundances in many clusters have shown that they cannot be described in such a simple way (see \citet{Gratton12} for a comprehensive review of these observations). Since there is no known process existing that can transport light elements from the core to the surface of a main sequence star, the abundance variations must have a primordial origin \citep{Hesser80}. These detections were complimented by the discovery of multiple distinct tracks along all sequences in the Hertzsprung-Russell diagrams of many globular clusters (e.g. \citet{Bedin04}; \citet{D'Antona05}; \citet{Milone08}; \citet{Piotto15}; \citet{Milone15}). The splitting of sequences in the HR diagram can be explained by populations of similar age and metallicity but different He abundances (e.g. \cite{Piotto07}). The group of stars with low He abundances is often referred to as the first population, while the He enriched population has been coined the second population. The second population is associated with many chemical peculiarities including anti-correlations between Na and O as well as Mg and Al \citep{Carretta09a,Carretta09b}. These observations have led to the conclusion that a globular cluster is made up of two (or more) populations that are chemically distinct from one another.

Several mechanisms for forming the second population of stars have been explored thus far. One main subset of theories works on the basic assumption that the second population forms out of material ejected by the first population. Due to the need for multiple star formation epochs, the populations are sometime referred to as multiple generations in this scenario. Several candidates for the polluting class of stars have been explored, including asymptotic giant branch (AGB) stars \citep{D'Ercole08} and fast rotating massive stars (FRMS) \citep{Decressin07}. \citet{D'Ercole08} used hydrodynamical and $N$-body simulations to study the evolution of material ejected from the first population in the AGB polluter scenario. They found that ejected material will collect in the centre of the initial cluster and the second generation will form concentrated in the central regions of the first generation. A general difficulty with this formation scenario, and many others, is the mass budget problem. Observations suggest that the second population can make up a significant fraction of the total present day population ($\approx$ 50 percent or higher) \citep{D'Antona08}, which is at odds with the amount of material available to form a second generation of stars. \citet{D'Ercole08} suggested this problem could be reconciled by the early loss of the majority of first generation stars. This loss would occur as the cluster expands past its tidal limit due to stellar evolution effects early in its life.

A second theory involves some fraction of cluster stars accreting polluted material early in their life and does not involve multiple epochs of star formation. One possible polluter studied in this context is interacting massive binary stars, first suggested by \citet{deMink09}. In this scenario, low-mass pre-main sequence stars accrete material via their circumstellar discs that has been enriched and ejected by massive interacting binaries \citep{Bastian13}. This model assumes that the cluster is mass segregated initially such that all ejected material is concentrated in the core. Low mass stars having orbits that bring them into the core ($\approx45$ per cent) will accrete polluted material and become a part of the second population while those stars that never enter the core remain pristine and make up the so called first population. The enriched population is initially more concentrated as a consequence of this mechanism. 

Although it is expected that the second population is more centrally concentrated initially, two body interactions will alter star orbits as the cluster relaxes and ultimately wash out the central concentration of the second population over time. This process is referred to as the spatial mixing of two populations. The mixing process has been studied previously using equal mass, low N models by \citet{Vesperini13}. They investigated the number of initial half mass relaxation times needed for mixing to occur for a variety of initial second to first population concentrations. They showed that second populations with higher central concentrations take more half mass relaxation times to mix than a population with a low central concentration. They also compared clusters with and without a point mass tidal field and found that the inclusion of a tidal field can significantly decrease the number of relaxation times needed to achieve mixing. This difference can be attributed to tidal stripping of the outer less mixed layers of the cluster which accelerates the mixing process. They also found that all of their clusters simulated in the tidal field were mixed when they had lost 60-70 per cent of their initial mass. Based on the number of present day relaxation times required for their simulated clusters to mix and typical relaxation times of globular clusters in the Milky Way, many Galactic globular clusters should not be mixed at present day and will retain some memory of the high initial concentration of the second population.

Previously, observational studies of multiple population clusters have found a second population which is more centrally concentrated than the first (e.g. \citet{Sollima07}; \citet{Bellini09};\citet{Lardo11}; \citet{Kravtsov11}; \citet{Beccari13}). However, \citet{Vanderbeke15} recently found that 80 per cent of the 48 Galactic clusters in their study have populations that are well mixed, often finding conflicting results for a large portion of clusters already studied. A possible explanation for these discrepancies is the method used for separating the cluster stars into first and second population. The study by \citet{Vanderbeke15} uses horizontal branch stars, while previous studies utilize the red giant branch to separate the populations.

The first discovery of fully mixed populations was made in the globular cluster NGC 6362 \citep{Dalessandro14}. The authors estimated a current half mass relaxation time of approximately 1.7 Gyr, noting that Galactic globular clusters with populations that are not mixed have similar relaxation times. The similarity in relaxation times indicates that something in addition to two body relaxation drove the mixing of the two populations in NGC 6362. Pointing towards the important role of mass loss in the mixing process, demonstrated by \citet{Vesperini13}, the authors suggested that this discrepancy could be reconciled if the cluster has lost a high fraction of its initial mass. They commented that the shallow present day mass function slope of NGC 6362, reported by \citet{Paust10}, is consistent with high amounts of mass loss due to tidal stripping, as shown by \citet{Webb14a}. The orbit of the cluster, found by \citet{Dinescu99} through the integration of proper motions, is inclined with respect to the plane of the Galaxy ($i = 22^{\degree}$) with small perigalactic ($R_p = 2.4$ kpc) and apogalactic ($R_a = 5.5$ kpc) distances. The high mass loss rates experienced on such an inclined and eccentric orbit \citet{Webb14a,Webb14} may be able to account for the full spatial mixing observed in this cluster, especially when a realistic Milky Way potential is considered (an aspect that has yet to be included in models of multiple population dynamics).

In this work, we perform simulations of clusters with multiple stellar populations on the same orbit as NGC 6362 in a Milky Way-like potential to test how the tidal field can affect the mixing process. We explore the effects of changing a variety of initial conditions, including the concentration of the second population, degree of tidal filling, cluster mass and fraction of second population stars.

\begin{table*}
\caption{Initial number of stars, concentration, half mass radii and half mass relaxation times for all simulations. Note: the size of the second population can be obtained by dividing the first population's size by the concentration, S. The half mass radius of the whole population is calculated after the two populations of different sizes are combined.}
\label{tab:parameters}
\begin{tabular}{| c | c | c | c | c | c | c | c | c |}
\hline
Name & Number of Stars & Concentration & \multicolumn{3}{c|}{Initial Half Mass Radius (pc)} & \multicolumn{3}{c|}{Initial Relaxation Time (Myr)} \\
\cline{4-9}
 & N & S & Total & First & Second & Total & First & Second \\
 \hline 
N50kS5 & 50,000 & 5 & 2.21 & 5.07 & 1.01 & 189.8 & 499.1 & 44.6 \\
N50kS5isol & 50,000 & 5 & 2.21 & 5.07 & 1.01 & 189.8 & 499.1 & 44.6\\
N50kS2.5 & 50,000 & 2.5 & 3.13 & 5.07 & 2.03 & 319.5 & 500.1 & 126.1\\
N50kS10 & 50,000 & 10 &1.48 & 5.07 & 0.51 &104.0 & 500.9 & 15.6 \\
N100kS5 & 100,000 & 5 & 2.22 & 5.17 & 1.01 & 252.5 & 677.0 & 58.1 \\
N100kS5R3 & 100,000 & 5 &1.33 & 3.10 & 0.60 & 117.4 & 314.6 & 27.0 \\
N50kS5R4 & 50,000 & 5 & 1.76 & 4.02 & 0.80 & 134.2 & 352.9 & 31.6 \\
N50kS5F70 & 50,000 & 5 & 3.19 & 5.08 & 1.00 & 535.7 & 626.2 & 23.1\\
\hline
\end{tabular}
\end{table*}

\section{Methods}

We follow similar methods to \citet{Vesperini13} to set up a cluster with two populations. We first generate two clusters independently using \textsc{mcluster} \citep{Kupper11}. These clusters follow \citet{King66} profiles with $W_0 = 7$, have an initial mass function as described by \citet{Kroupa01}, with masses in between 0.1 and 50 $M_{\odot}$ and an average mass of 0.6 $M_{\odot}$, and are initially virialized. When the clusters are initially generated they have the same size. One cluster is then scaled down to represent the more centrally concentrated second population. This cluster is scaled such that the ratio of the half mass radius (radius that contains half the mass) of the first population ($R_{FP}$) to the half mass radius of the second population ($R_{SP}$), S, takes on the desired value. The scaled cluster's velocities are adjusted such that it remains virialized with respect to itself. Finally the two clusters are combined to make one, at which time all the velocities of both populations are scaled such that the entire combined cluster is virialized. These clusters are then input into the \textsc{Nbody6} code \citep{Aarseth01,Aarseth03}. For all simulations, the effects of stellar evolution are implemented in the code.

We simulate the clusters in a galactic potential representative of the Milky Way. The potential is made up of a point mass bulge, disc \citep{Miyamoto75} and logarithmic halo. Details of this model and values of the parameters used can be found in \citet{Miholics14}. To obtain the orbit of NGC 6362, we simply integrate the proper motions (taken from \citet{Dinescu97}) of the cluster from its current position in the Milky Way, (X,Y,Z) = (2.0 kpc, -4.1 kpc, -2.3 kpc) \citep{Harris96} (corresponding to an initial Galactocentric distance of 5.1 kpc). We obtain a similar orbit for the cluster as found by \citet{Dinescu99} ($R_a = 5.5$ kpc, $R_p = 2.4$ kpc, $i = 22^{\degree}$), although not exactly the same due to the differences in galactic potentials used.

We use a simulation of a 50,000 star cluster consisting of two populations equal in number as our base simulation to which we compare all our models. The first population has an initial half mass radius of 5 pc. The ratio of first population half mass radius to second population half mass radius in the base simulation is 5 ($S = 5$). The number of stars in each population was chosen to be equal to roughly emulate the current observed fraction of first population stars in most globular clusters, $\approx50$ percent or lower \citep{D'Antona08}. The base simulation will be referred to as N50kS5. To study the effect of the Milky Way potential, we simulate the same cluster in isolation, N50kS5isol. We also examine the consequences of changing the initial concentration of the second population by performing simulations with S= 2.5 and S =10, N50kS2.5 and N50kS10 respectively. We also simulate two $N=100,000$, $S=5$ clusters with different first population sizes, 5 pc and 3 pc. These simulations will be referred to as N100kS5 and N100kS5R3. The truncation radii of these clusters are 43.6 pc and 26.1 pc respectively. Changing the size of the cluster, changes the degree to which it fills it tidal (or Jacobi radius), $r_j$, which is, roughly speaking, the radius at which a cluster star feels the same force from the cluster as it does from the galaxy. The initial Jacobi radius of these clusters is the same since they have the same mass and initial position in the Milky Way and is calculated to be 21.1 pc \citep{Bertin08}. The degree to which a cluster fills its tidal radius is often called the tidal filling factor and is calculated as the ratio of half mass radius ($r_h$) to Jacobi radius ($r_j$) (e.g. \citet{Baumgardt08}). The initial tidal filling factors of these two clusters are 0.11 and 0.07 respectively. We note that as a result of the small Jacobi radius on the adopted orbit, these clusters have truncation radii larger than the Jacobi radius. However, due to the smaller second population, the cluster is quite centrally concentrated, with a large majority of stars lying well inside the Jacobi radius (the radius containing 95 per cent of the cluster's mass is 16.05 pc and 9.63 pc, for N100kS5 and N100kS5R3 respecitvely). In fact, these clusters are still considered underfilling initially since their filling factors are less than 0.145. \citep{Henon61,Alexander13}. 

Comparing simulations with different N allow us to see how the results scale with N and extend our results to NGC 6362. The shallow mass function slope of NGC 6362 $\alpha = -0.49$ \citep{Paust10} suggests that the cluster has lost a significant amount of mass over its lifetime \citep{Vesperini97,Webb14a}. We can estimate the initial mass of this cluster by utilizing the results of \citet{Webb15}, who demonstrated the following relationship between fraction of mass lost and mass function slope:
\begin{equation}
\alpha = 2.8e^{-6.4\frac{M}{M_{i}}} - 1.5
\end{equation}
where M is the current mass and $M_i$ is the initial mass. This relationship was found by fitting simulations of clusters with a wide range of initial sizes, masses and orbits and shows minimal scatter. Using a present day mass of $5 \times 10^4 M_{\odot}$ \citep{Dalessandro14} and this relation, we calculate that NGC 6362 has lost $\approx$ 85 per cent of its mass, giving it an initial mass of $\approx 3.3 \times 10^5 M_{\odot}$. A cluster of this size corresponds to a simulation of over 500,000 stars, which is too computationally expensive to realistically perform.

Finally, we simulate an $N = 50,000$ star cluster similar to the base cluster but with 70 per cent first population stars, N50kS5F70. Altering this fraction allows us to study how different initial fractions of first population stars affects the fraction of first population stars at the time of mixing. Simulations are run until the cluster contains only 100 stars, i.e. until the cluster has dissolved. All simulations with initial parameters, and relaxation times, are listed in Table \ref{tab:parameters}.
\begin{table*}
\caption{Some parameters of interest evaluated at the time of mixing for all simulations: time ($t_{mix}$), number of relaxation times ($n_{mix}$), fraction of mass lost ($m_{loss}$), fraction of second population stars ($f_2$).}
\label{tab:results}
\begin{tabular}{| c | c | c | c | c |}
\hline
Name & $t_{mix}$ (Myr) & $n_{mix}$ & $m_{loss}$ & $f_2$ \\
 \hline 
N50kS5 & 2160 & 11.4 & 0.71 & 0.64 \\
N50kS5isol & N/A & N/A & N/A & N/A\\
N50kS2.5 & 1510 & 4.7 & 0.66 & 0.59 \\
N50kS10 & 2020 & 19.4 & 0.73 & 0.67 \\
N100kS5 & 4760 & 18.8 & 0.79 & 0.64 \\
N100kS5R3 & 4610 & 39.3 & 0.76 & 0.58 \\
N50kS5R4 & 2500 & 18.5 & 0.75 & 0.62 \\
N50kS5F70 & 2240 & 6.8 & 0.78 & 0.41\\
\hline
\end{tabular}
\end{table*}

\section{Results}

To quantify when the populations are considered spatially mixed, we select the first time-step where their half mass radii differ by less than 0.01 pc. In Table \ref{tab:results}, we have listed several parameters evaluated at mixing for all simulations, including the time of mixing, number of relaxation times, fraction of mass lost (including mass loss from stellar evolution), and fraction of second population compared to the current total. These parameters help to characterize whether or not a cluster should be spatially mixed at present day.

\subsection{Effects of Realistic Milky Way Potential}

\begin{figure} 
\begin{subfigure}[b]{0.5\textwidth}
\centering
\includegraphics[width=.8\linewidth]{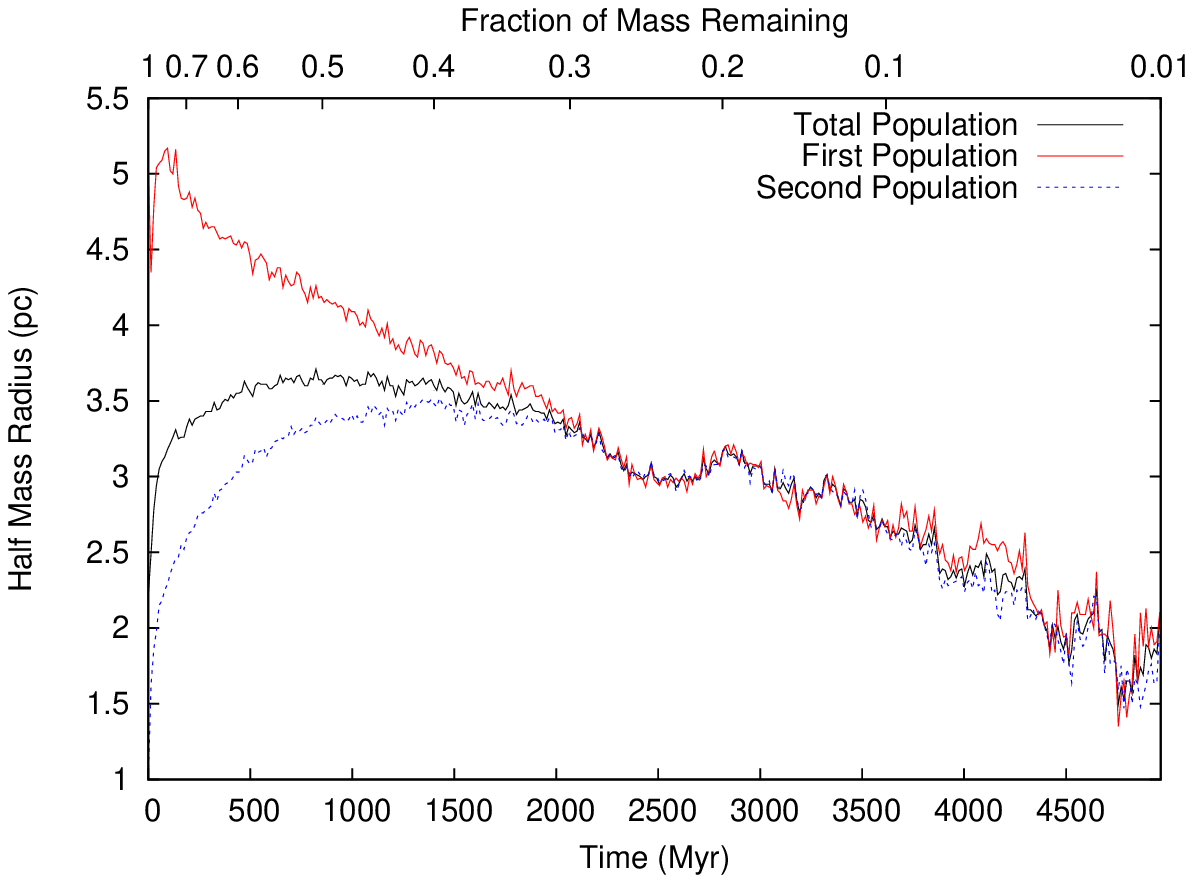}
\caption{N50kS5}
\label{fig:Base}
\end{subfigure}
\begin{subfigure}{.5\textwidth}
\centering
\includegraphics[width=.8\linewidth]{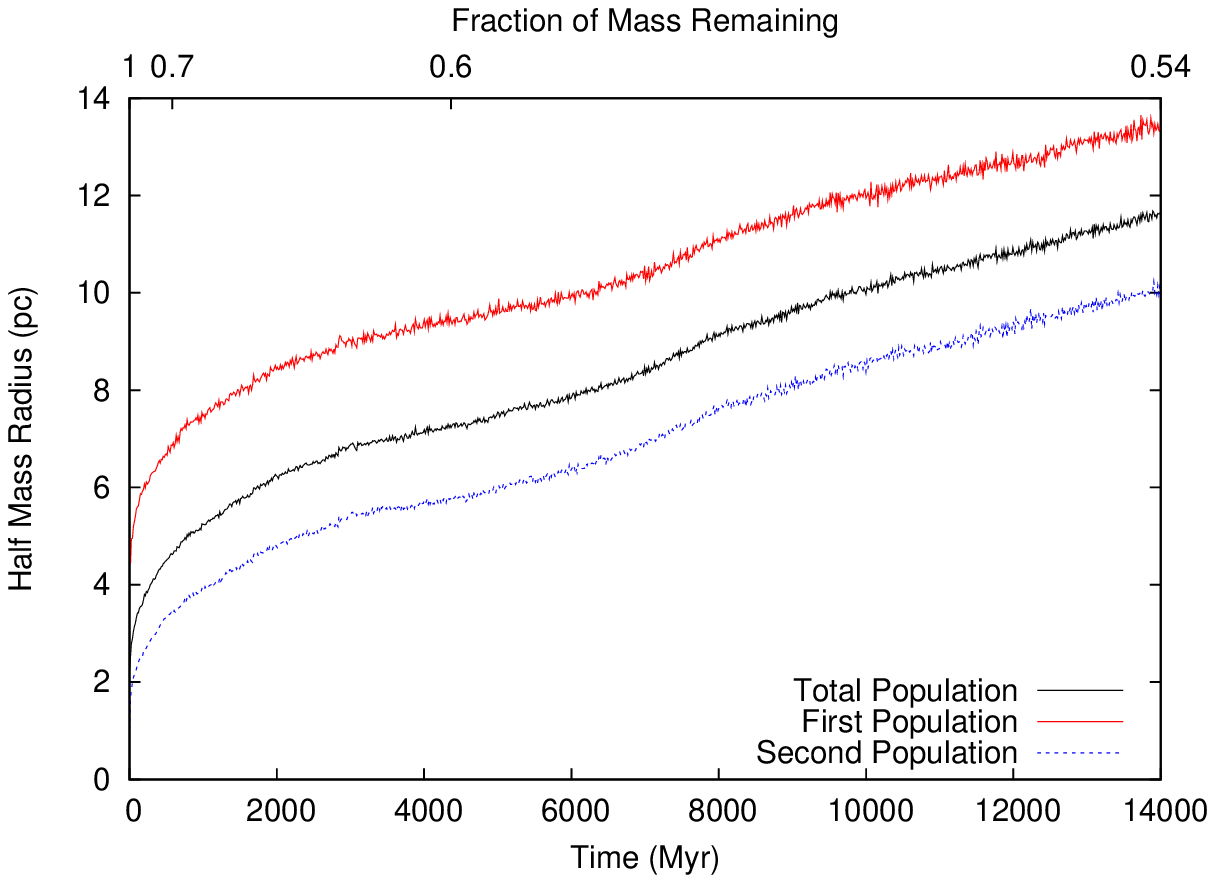}
\caption{N50kS5isol}
\label{fig:Isolation}
\end{subfigure}
\caption{Half mass radii of total (black), first (red) and second (blue) populations against time and ratio of current mass to initial mass (upper axis)} for two $N=50,000$, $S=5$ clusters, one evolved on the orbit of NGC 6362 in the Milky Way (base simulation) and one in isolation.
\label{fig:MW}
\end{figure}

The first panel of Figure \ref{fig:MW} depicts the half mass radii of the first, second and total population (represented by red, blue and black lines respectively) of the base simulation, N50kS5. The two populations are spatially mixed after 2160 Myr or approximately 11.4 relaxation times ($t_{mix}$ and $n_{mix}$ in Table \ref{tab:results}). A sharp decrease in the initial size of the first population is observed in the first 2 Gyr. Since the tidal field on the cluster is strong within the inner parts of the Galaxy, the outer portions of the cluster, consisting primarily of first population stars, are stripped away quickly by the Galaxy. This preferential stripping allows for complete spatial mixing of the two populations to occur early in the cluster's life, at less than half of its dissolution time. This evolution is in contrast to the behaviour of the same cluster when evolved in isolation, depicted in the second panel of Figure \ref{fig:MW}. The cluster evolved in isolation loses very little mass compared to the cluster on the orbit of NGC 6362 and there is no strong preferential loss of first population stars. The populations expand in step with one another and despite many relaxation times of evolution the isolated cluster maintains a strong memory of the initial second population concentration.

\subsection{Effect of Second Population Concentration} 

\begin{figure}
\begin{subfigure}[b]{0.5\textwidth}
\centering
\includegraphics[width=.8\linewidth]{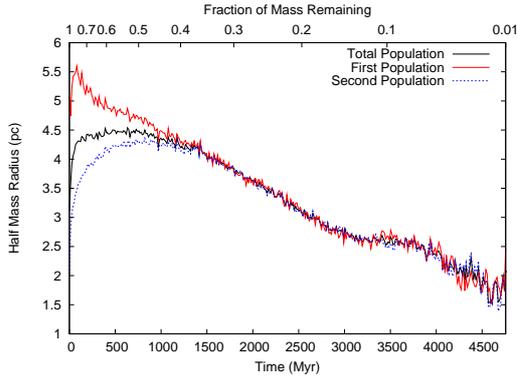}
\caption{N50kS2.5}
\label{fig:Concentrations2.5}
\end{subfigure}
\begin{subfigure}{.5\textwidth}
\centering
\includegraphics[width=.8\linewidth]{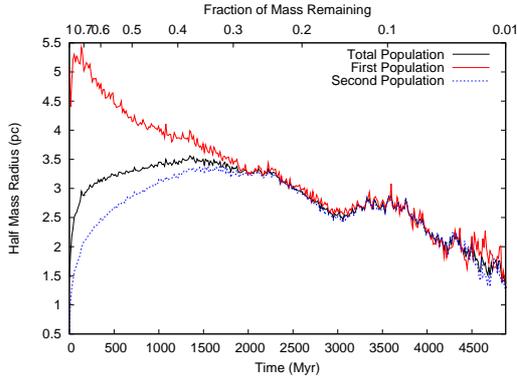}
\caption{N50kS10}
\label{fig:ConcentrationS10}
\end{subfigure}
\caption{Half mass radii of total, first and second populations (colours the same as Figure \ref{fig:MW}) for two clusters similar to the base simulation but with $S=2.5$ and $S=10$ respectively.}
\label{fig:Concentrations}
\end{figure}

The half mass radii of clusters in simulations N50kS2.5 and N50kS10 are depicted in Figure \ref{fig:Concentrations}. These clusters were obtained by keeping the half mass size of the first population the same (5 pc) and scaling the second population differently. N50kS2.5 has a second population twice as large as N50kS5 while N50kS10 has a second population twice as small.

The number of relaxation times at mixing becomes larger as the concentration of the second population with respect to the first population is increased. This effect results from the increased amount of structure that must be washed away before complete mixing can be achieved. For higher concentrations, the cluster is also less susceptible to tidal stripping due to decreased half mass radius. These simulations show that, without knowledge of the initial concentration, the number of relaxation times that have passed does not characterize whether the cluster should be mixed. However, on inspection of the parameters in Table \ref{tab:results}, we see that many similarities between the clusters exist. For example, at mixing, the clusters have similar fractions of mass lost, in the range of 65 to 75 per cent. The number ratio of second to first population stars also appears to be a good indication of spatial mixing, as the clusters are all mixed when the fraction of second population stars is $\approx$ 60-70 per cent of the cluster. This fraction is dependent on the initial number ratio of first to second population stars, as will be discussed in Section 3.5. All three of these clusters dissolve after about 5 Gyr. Spatial mixing in each simulation occurs early within the cluster's life, well before half the dissolution time.
 
\subsection{Dependance on Tidal Filling Factor}

\begin{figure}
\begin{subfigure}{.5\textwidth}
\centering
\includegraphics[width=.8\linewidth]{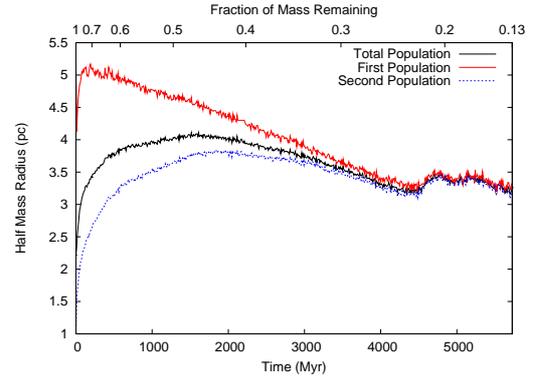}
\caption{N100kS5}
\label{fig:100kbig}
\end{subfigure}
\begin{subfigure}{0.5\textwidth}
\centering
\includegraphics[width=.8\linewidth]{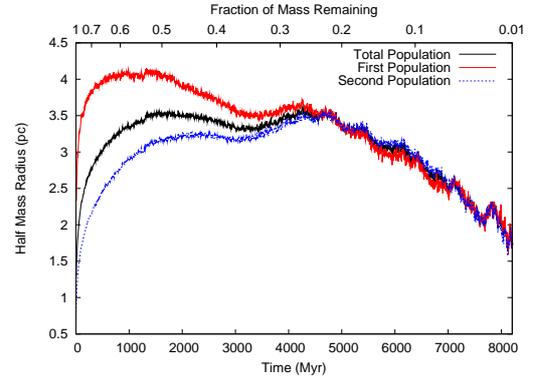}
\caption{N100kS5R3}
\label{fig:100ksmall}
\end{subfigure}
\caption{Half mass radii for total, first and second population (colours the same as Figure \ref{fig:MW}) of two $N=100,000$ star clusters with different initial sizes. The top panel represents a cluster with initial first population size 5 pc and bottom panel represents a cluster with initial first population size 3 pc.}
\label{fig:TidalFilling}
\end{figure}

We examine the effect of changing the size, and thus the tidal filling factor ($r_h/r_j$) (see Section 2 for details), of the cluster by performing two $N = 100,000$ simulations with initial first population half mass radii of 5 pc and 3 pc (N100kS5 and N100kS5R3). Initially, N100kS5 has a tidal filling factor of 0.11 while N100kS5R3 has a filling factor of 0.07. Figure \ref{fig:TidalFilling} shows the evolution of half mass radii for these two clusters. Despite being underfilling initially, N100kS5R3 expands quickly and is tidally filling for most of its life. This effect is a consequence of the strong tidal field experienced by the cluster on its orbit in the inner regions of the Galaxy. 

The physical time of mixing for these clusters is very similar, a consequence of two competing effects. The cluster in N100kS5R3 has a shorter relaxation time since it is smaller and we would therefore expect this cluster would mix faster than its larger counterpart. However, the rate of mixing is accelerated for N100kS5 since it is more tidally filling. The outer layers of this cluster are more susceptible to being stripped by the Galaxy and many first population stars are lost early. Hence, N100kS5 needs fewer relaxation times to mix. At the time of mixing, the clusters have experienced similar amounts of mass loss and have similar fractions of first and second population stars (refer to Table 2).

\subsection{Dependance on Cluster Mass}

In order to compare our simulations with the real cluster NGC 6362, we must know how our results depend on the number of stars in the cluster. To this end, we compare the base simulation with N100kS5, a simulation with twice the stars but the same initial size. We also simulate a new 50,000 star cluster with the same tidal filling factor as N100kS5, N50kS5R4. Hence, we can compare $N = 50,000 $ and $N =100,000$ star clusters that have the same susceptibility to mass loss due to tidal stripping. 

\begin{figure}
\centering
\includegraphics[width=0.8\linewidth]{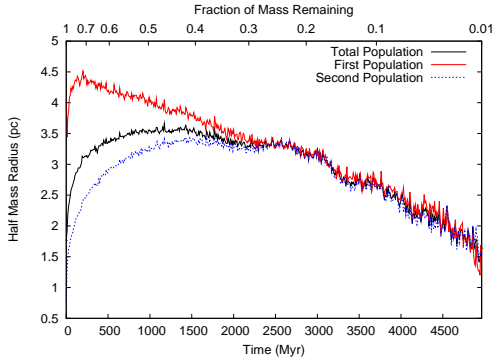}
%\caption{N50kS5R4}
%\label{fig:50ksmall}
\caption{Half mass radii for total, first and second population (colours the same as Figure \ref{fig:MW}) of N50kS5R4, a $N=50,000$ star cluster with the same tidal filling factor as N100kS5 (shown in Figure \ref{fig:100kbig}).}
\label{fig:Mass}
\end{figure}

When comparing the base simulation (Figure \ref{fig:Base}) with N100kS5 (Figure \ref{fig:100kbig}), we see N100kS5 takes about twice as long to mix, despite similar relaxation times. The primary difference between these clusters are their tidal radii; the more massive cluster will have a larger tidal radius. Since the half mass radii are the same, the more massive cluster is less tidally filling and experiences slower mass loss rates per unit mass. At the time of mixing, the fractions of mass lost are comparable: 79 and 71 per cent for $N=100,000$ and $N=50,000$ respectively. However, the strongest similarity between the two clusters comes from examining the fractions of first and second population stars at the time of mixing. These values are identical, each cluster being made up of approximately 36 per cent first population (64 per cent second population) stars. The evolution of the $N = 100,000$ cluster is even more similar to the $N = 50,000$ with the same tidal filling factor (N50kS5R4), shown in Figure \ref{fig:Mass}. These clusters are essentially scaled versions of one another within the framework of the tidal field and their mass loss rates per unit mass will be similar. This phenomenon is reflected in their almost identical number of relaxation times until mixing and fraction of mass lost at time of mixing. The fraction of first population stars at mixing for N50kS5R4 is 38 per cent, similar to the fraction for N100kS5 of 36 per cent as stated above. The similarities between these clusters clearly demonstrates the link between mass loss and spatial mixing. Hence, for a cluster of higher initial mass, the time for spatial mixing can be estimated from fraction of mass lost at mixing found from simulations of its low N counterparts.

\subsection{Dependance on Initial Number of First and Second Population Stars}

\begin{figure}
\centering
\includegraphics[width =0.8\linewidth]{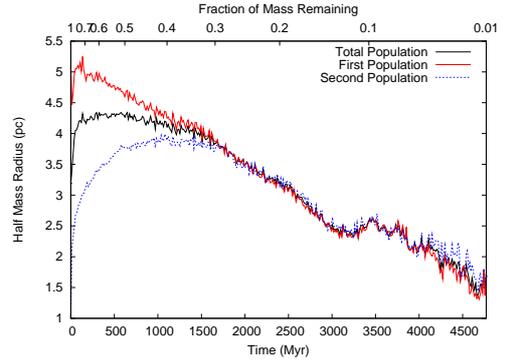}
\caption{Half mass radii for total, first and second population (colours the same as Figure \ref{fig:MW}) of N50kS5F70, a cluster starting with 70 per cent first population stars.}
\label{fig:Fraction}
\end{figure}

We explore the dependance of our results on the initial ratio of first and second populations stars by running a simulation with 70 per cent first population stars, N50kS5F70. The initial half mass radius of the first population is 5 pc and $S = 5$, similar to the base simulation. The results are illustrated in Figure \ref{fig:Fraction}. Mixing occurs at approximately the same physical time for N50kS5 and N50kS5F70. The cluster in N50kS5F70 has a longer relaxation time which would tend to slow the mixing process. However, this effect is compensated for by the cluster's slightly larger half mass radius (calculated for the combined populations) over the base cluster. The increased size means that the cluster's less mixed outer layers are more vulnerable to tidal stripping and this balances the effect of a long relaxation time. The cluster achieves mixing when 78 per cent of the initial cluster mass has been lost, similar to other simulations in our study. However, the higher fraction of first population stars leaves a distinct imprint on the cluster. The most notable difference between N50kS5F70 and previous simulations is the fraction of first and second population stars at mixing. Most of the clusters in our study are mixed when the first population is about 30-40 per cent of the cluster. The fraction of first population stars in this cluster is 59 per cent at mixing, a reflection of the higher initial fraction of first population. The cluster retains a memory of its initial fraction of first population stars regardless of the tidal field strength. The expansion due to stellar evolution is not enough to preferentially remove most first population stars even when the tidal field is strong.

\section{Summary and Discussion}

We have shown that the mixing of multiple populations in globular clusters depends strongly on the tidal field in which the cluster lives. The differences in multiple population dynamics of clusters embedded in a point mass tidal field and isolation were first noted by \citet{Vesperini13}, who found that tidal stripping significantly accelerates the spatial mixing process. In general, the number of relaxation times required for full mixing in the point mass tidal field is not the same for a cluster in the Milky Way since clusters lose mass at different rates. This reinforces the findings of \citet{Vesperini13} and in particular shows that the rate of mass loss can have a large effect on the physical time of mixing. Mass loss rates for Milky Way clusters will in general vary according to Galactocentric distance, eccentricity and inclination\citep{Baumgardt03,Webb14a,Webb14}. Hence, the mixing of multiple populations in Milky Way globular clusters will depend strongly on a cluster's orbit. Clusters close in to the centre of the Galaxy with high inclinations, such as NGC 6362 studied here, are much more likely to be mixed at present day than clusters with high mean Galactocentric distances. 

We have also shown how altering different parameters such as initial subpopulation concentration, tidal filling factor and number of stars alters the spatial mixing of two populations. In general, we have shown that altering any one of these parameters does not significantly alter the amount of mass loss required for spatial mixing. Additionally, all of these models have similar fractions of second population stars at the time of mixing. Any differences between models in terms of fraction of mass lost, fraction of second population, time for mixing, and number of initial relaxation times for mixing can in general be explained considering two competing effects: mixing driven by two body interactions and mixing due to tidal stripping. At the beginning of each cluster's evolution, the mass that is stripped is primarily made up of first population stars since they initially dominate the outer regions. Mass loss rates will be slower initially for tidally underfilling clusters, allowing relaxation processes to start mixing outer layers of the cluster before they are stripped. Hence, less tidally filling clusters retain slightly more first population stars and lose less mass by the time mixing is achieved. Conversely, clusters with higher second population concentrations need to lose a larger fraction of their mass, and a higher fraction of first population stars, before mixing since a higher fraction of their volume is dominated by first population stars. The cluster's size, and thus mass, decrease significantly before the second population's size becomes comparable to the first.

Finally, by altering the initial fraction of first population stars we have demonstrated that that this fraction does not significantly alter the dynamics of the two populations. The fraction of mass lost at the time of mixing for N50kS5F70 is similar to the rest of our models. The biggest imprint of a higher initial fraction of first population stars is the resulting fraction of first population stars at mixing. Our other simulations have shown that if this cluster's second population is more centrally concentrated and/or the cluster was born more tidally filling, than the fraction of first population stars at mixing will be lower. However, for clusters with similar second population concentrations and filling factors, we have shown that a higher initial fraction of first population stars should in general lead to a higher fraction of first population stars at mixing.

The universal indicator of when our model clusters become mixed is fraction of initial mass lost; most clusters in our study are mixed when the cluster has lost 70-80 per cent of its initial mass. The fraction of mass lost by NGC 6362 is estimated to be approximately 85 per cent, based on its shallow present day mass function \citep{Webb15}. Hence, the observed complete spatial mixing of the two populations in this cluster is fully consistent with the results from our models.

Our results allow us to comment on the initial fraction and concentration of second population stars in NGC 6362. At present day, NGC 6362 contains approximately 55 per cent second population stars \citep{Dalessandro14}. We argue that this is representative of the fraction of second population stars at the time of mixing since clusters should lose first and second population stars at an approximately equal rate after mixing. To say something meaningful about initial fraction and concentration of second population, we must know if the cluster was tidally filling or underfilling at birth. The cluster is tidally filling at present day, with an observed filling factor (ratio of half light radius to Jacobi radius) of $r_{hl}/r_j = 0.23$ (where $r_{hl}$ was taken from \citet{Harris96} and the method of \citet{Bertin08} was used to calculate $r_j$). Since the cluster is currently close to apogalacticon, it is likely filling along its entire orbit. We argue that this has been the case for most of the cluster's life since our models indicate that on this orbit clusters become tidally filling quickly, regardless of how underfilling they are at birth. Under this assumption, it is most likely that the cluster was born with a small fraction of second population stars with a high concentration. In this case, the cluster would experience a large preferential tidal stripping of first population stars over its lifetime leaving the cluster with a roughly equal number of first and second population stars present day. 

Since the mixing process is so closely tied to mass loss, our results can be easily applied to other Milky Way clusters on different orbits. \citet{Dalessandro14} estimated a present day relaxation time for NGC 6362 of 1.7 Gyr and noted that this was similar to relaxation times of globular clusters with populations not spatially mixed at present day. Our models indicate that this discrepancy can be easily rectified if non-mixed clusters have lost a lower fraction of their mass than NGC 6362 (less than 85 per cent). This idea is supported by the four Galactic globular clusters, NGC 5272, 5904, 6205, 6341, which have measured mass function slopes as well as known radial distributions for their first and second populations. All four of these clusters have been shown to be not spatially mixed at present day \citep{Lardo11} and have steeper mass functions than NGC 6362 \citep{Paust10}, corresponding to lower fractions of mass lost \citep{Webb15}. These clusters have lost in between 70-80 per cent of their mass, supporting the idea that the second population consists of a small number of stars highly concentrated in the central regions of the cluster. Under these conditions a cluster would need to lose a higher fraction of mass than the models presented here to become mixed. 

We have shown that mass loss due to tidal stripping, or otherwise, in the Milky Way is essential to the mixing of multiple populations in globular clusters. Hence, when considering whether a Galactic globular cluster should be mixed by present day a simple estimate of relaxation time is not a sufficient indicator. It is important to consider the cluster's orbit in the Milky Way and how much mass the cluster has lost over its lifetime. Furthermore, we have shown that it is possible to use measurements of fraction of mass lost for clusters that are not yet spatially mixed to place constraints on currently unknown parameters, such as initial second population concentration and number of stars in each population.

\section*{Acknowledgments}
We acknowledge the use of Sharcnet resources in the completion of this work. A.S. and J.W. are supported by NSERC (Natural Sciences and Engineering Research Council of Canada).

\label{lastpage}

\end{document}